\documentstyle[11pt,paspconf,twoside,epsf]{article}

\begin{document}

\title{Cluster environments around quasars at $0.5 \leq z \leq 0.8$}
\author{M. Wold$^{1}$, M. Lacy$^{2}$, P.B. Lilje$^{3}$, S. Serjeant$^{4}$}
\affil{$^{1}$ Stockholm Observatory, SE-133 36 Saltsj{\"o}baden, Sweden}
\affil{$^{2}$ IGPP, L-413, LLNL, Livermore CA 94550, USA}
\affil{$^{3}$ Institute of Theoretical Astrophysics, University of Oslo, 
P.O. Box 1029 Blindern, N-0315 Oslo, Norway}
\affil{$^{4}$ Astrophysics Group, Imperial College London, Blackett 
Laboratory, Prince Consort Road, London SW7 2BZ, U.K.}

\begin{abstract}
We have observed the galaxy environments around two complete samples of 
radio-loud (steep-spectrum) and radio-quiet quasars (RLQ and RQQ)
at $0.5 \leq z \leq 0.8$ that are matched in $B$-luminosity, and find that 
the environments of both quasar populations are practically indistinguishable. 
A few objects are found in relatively rich clusters, but on average, they 
seem to prefer galaxy groups or clusters of $\approx$ Abell class 0.

By combining the RLQ sample with samples from the literature, 
we detect a weak, but significant, positive correlation between environmental
richness and quasar radio luminosity. This may give us clues about what 
determines a quasar's radio luminosity.
\end{abstract}


\section{Introduction}

Powerful AGN can be used as tracers of galaxy groups and clusters.
Especially, radio-loud AGN at $z\sim0.6$ are known to lie in 
rich environments (Ellingson, Yee, \& Green 1991; Yee \& Green 1987).
However, the environments of RQQ's have been found to be systematically
different from RLQ environments, suggesting that they prefer
field-like environments (Ellingson et al.\ 1991).
Recent developments are beginning to cast doubt on this, e.g\ 
McLure et al.\ (1999) find that luminous RQQ's have similar massive
elliptical host galaxies as their radio-loud counterparts.

We have carried out a study of the environments of RLQ's and RQQ's
at $0.5 \leq z \leq 0.8$ with the 2.56m Nordic Optical Telescope and the HST.
The radio-loud sample consists of steep-spectrum quasars 
spanning a wide range in radio luminosity,
and the radio-quiet sample matches the radio-loud sample in $B$ luminosity.
By covering a wide range in luminosity and a narrow redshift range,
we aim to disentangle the effects that redshift and luminosity have
on the amount of clustering around quasars.

The excess of galaxies in each quasar field was obtained by subtracting 
background galaxy counts in observed control fields, and the clustering 
quantified as $B_{\rm gq}$, the amplitude of the spatial 
galaxy-quasar cross-correlation function.

\section{Results}

\begin{figure}
\plottwo{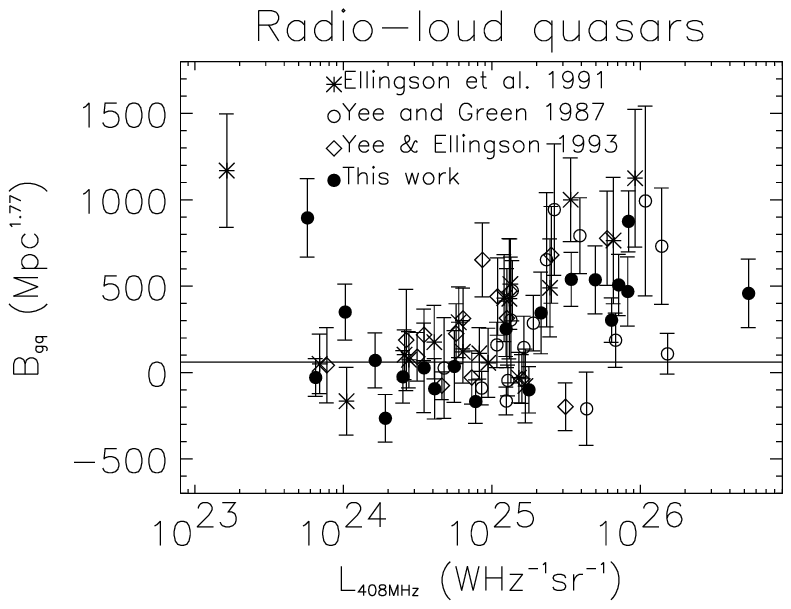}{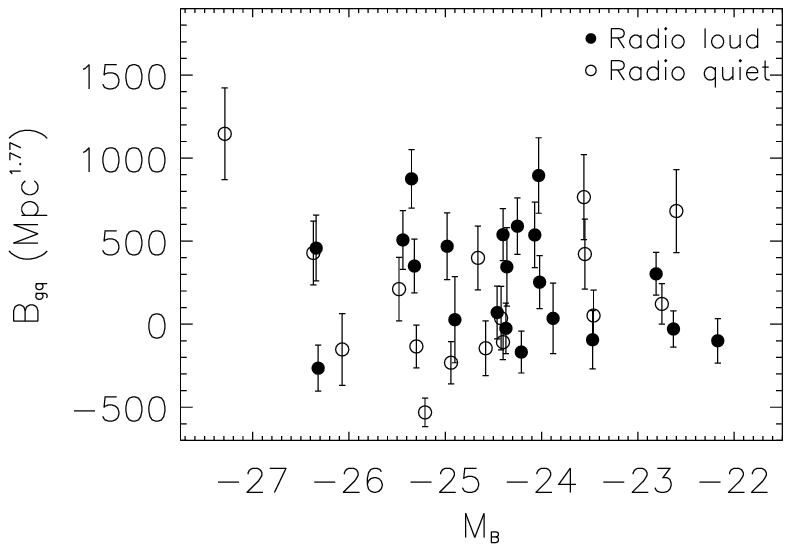}
\caption{Left: $B_{\rm gq}$ for our sources + literature sources
as a function $L_{\rm 408 MHz}$. Right: $B_{\rm gq}$ as a function of $M_{B}$ for 
RLQ's and RQQ's.}
\label{fig:figure1}
\end{figure}

On average, the quasars were found to occupy environments
typical of poorer clusters ($\approx$ Abell 0), but the richness was seen to
cover a wide range, from groups of galaxies and poor clusters to 
clusters as rich as Abell class 1 or more (Wold et al.\ 1999).
The mean clustering amplitude
for the radio-loud sample was found to be $B_{\rm gq}=265$ Mpc$^{1.77}$ 
with an error in the mean of 74 Mpc$^{1.77}$ 
(from a combination of intrinsic dispersion in $B_{\rm gq}$ and measurement error).
Preliminary analysis of the radio-quiet sample
shows that RQQ's are found in similar environments
as the RLQ's (Fig.~\ref{fig:figure1}, right), the two samples have practically 
the same mean richness. This is contrary to what has been observed before 
(Ellingson et al.\ 1991), but broadly consistent with luminous RQQ's having
similar host properties as RLQ's (McLure et al.\ 1999).

Using Spearmans partial rank correlation analysis, we find
a weak, but significant correlation between $B_{\rm gq}$ and quasar 
radio luminosity, $L_{\rm 408 MHz}$, holding redshift constant. 
This correlation has a 3.4$\sigma$ significance when we add
$B_{\rm gq}$ for literature steep-spectrum quasars extending down
to $z\approx0.2$ (Fig.~\ref{fig:figure1}, left). 
We find no evidence for any $z$-dependence in richness. 

In Wold et al. (1999), we compare our data with a simple radio source model, 
and find that the observed range in environmental density cannot account 
on its own for the range in radio luminosity, implying that the
bulk kinetic power in the radio jets must also be an important factor
in determining a quasar's radio luminosity.

\end{document}